# Transparency in Healthcare AI:
# Testing European Regulatory Provisions against Users' Transparency Needs

**Anna Spagnolli[1,3], Cecilia Tolomini[2], Elisa Beretta[1], Claudio Sarra[2,3]**

[1]Dipartimento di Psicologia generale, Università degli Studi di Padova

[2]Dipartimento di Diritto Privato e Critica del Diritto, Università degli Studi di Padova

[3]Human Inspired Technologies Research Centre, Università degli Studi di Padova

**Abstract**

Artificial Intelligence (AI) plays an essential role in healthcare and is pervasively incorporated into medical software and equipment. In the European Union, healthcare is a high-risk application domain for AI, and providers must prepare Instructions for Use (IFU) according to the European regulation 2024/1689 (AI Act). To this regulation, the principle of transparency is cardinal and requires the IFU to be clear and relevant to the users. This study tests whether these latter requirements are satisfied by the IFU structure. A survey was administered online via the Qualtrics platform to four types of direct stakeholders, i.e., managers (N = 238), healthcare professionals (N = 115), patients (N = 229), and Information Technology experts (N = 230). The participants rated the relevance of a set of transparency needs and indicated the IFU section addressing them. The results reveal differentiated priorities across stakeholders and a troubled mapping of transparency needs onto the IFU structure. Recommendations to build a locally meaningful IFU are derived.

**Keywords:** transparency, AI Act, healthcare, user-centeredness

## 1. Introduction

The software called Artificial Intelligence is the object of recent regulations and guidelines such as the European Union AI Act (Artificial Intelligence Act, 2024), the US AI Risk Management Framework (NIST USA, n.d.), or UNESCO's recommendations on the Ethics of Artificial Intelligence (UNESCO, 2022). Overall, these initiatives aim to increase the trustworthiness of AI technology, especially in application domains where mistakes and misuse have high costs for human well-being and rights. According to European law, health applications represent one such domain.

To minimize these risks, the AI Act prescribes that providers make available to users (or "deployers," in the regulation terminology) some Instructions for Use (IFU) about the systems' capabilities, limitations, and security. These instructions implement the obligation to transparency, facilitating an informed, responsible, and proper use of high-risk AI technology. Full compliance with the principle of transparency also requires that such information is meaningful to the actual users (e.g., Ada Lovelace Institute et al., 2021; Caitlin Vogus, Emma Llansó, 2021). In other words, some care must be taken to design the Instructions for Use so that they are understandable and relevant to users.

The process leading to the AI Act was participatory, and the resulting text was iteratively shared and debated with stakeholders since 2018[1]. Then, following the enforcement of the regulation in 2024, a reverse process has started to implement the regulation in each application. For the EU formal bodies, implementation includes holding public consultations to draft a Code of Practice for general-purpose AI or organizing best practice workshops with providers' and deployers' organizations. For most providers, implementation means – among other things - finding a way to prepare

---

Corresponding author: Anna Spagnolli, Dip. Di Psicologia Generale, via Venezia 8, 35100 Padova, Italy (anna.spagnolli@unipd.it)

[1] https://digital-strategy.ec.europa.eu/en/policies/european-approach-artificial-intelligence

the needed documentation and inform users, hopefully joining legal and user-centered transparency. This study aims to contribute to this local process, focusing on the structure of the IFU as proposed in the regulation.

We assessed whether the IFU structure represents an effective entry point to transparency information or needs to be refined before being presented to stakeholders. To our knowledge, no previous study has attempted to do so. Gils' team (Gils et al., 2024) carried out design workshops with 21 and 15 stakeholders; their sample is insightful yet small and considers several application scenarios besides healthcare. In our study, we collected transparency needs from the literature and involved 812 participants distributed across four categories of stakeholders, i.e., managers, health professionals, patients, and IT experts.

The paper is organized as follows. First, we contextualize the object of investigation by defining the domain of healthcare AI and its related stakeholders. Then, we introduce the concept of transparency and transparency needs. We then describe the study method and summarize the results with descriptive statistics. The results are then discussed and translated into design recommendations. The contributions of this study consist of (a) testing the IFU's structure against common transparency needs in healthcare AI; (b) providing design recommendations to improve its responsiveness to such needs; (c) applying an evaluative approach that can be exported to other application domains or more specific AI systems.

## 2. AI in healthcare

### 2.1 Applications

Artificial Intelligence can analyze massive amounts of data to make inferences, recognize patterns, and draw predictions. In healthcare, these capabilities are applied to diagnostic predictions, knowledge organization, decision support, surgical assistance, and remote protocols, as described by recent literature reviews (Al Kuwaiti et al., 2023); (Azzi et al., 2020) (Chellasamy & Nagarathinam, 2022); (Chustecki, 2024) (Kitsios et al., 2023) (Secinaro et al., 2021) (Sharma & Jindal, 2023) (Singh et al., 2023)(Wang et al., 2024))

AI can assist the physician in clinical practice by combining the patients' medical history, symptoms, and test results (Kitsios et al., 2023). Based on this data, diagnostic AI can make predictions (e.g., health risks) at the individual level and support the interpretation of medical images (Sharma & Jindal, 2023)(Kumar et al., 2023) (Al Kuwaiti et al., 2023); (Secinaro et al., 2021) (Singh et al., 2023); (Wang et al., 2024) (Chustecki, 2024)(Wang et al., 2024); (Chustecki M., 2024). AI solutions can organize a knowledge domain for patients and health professionals; the former can receive recommendations or obtain scientifically valid information about medical conditions and pathologies (Azzi et al., 2020), the latter can find scientific articles and updated reports (Al Kuwaiti et al., 2023). AI systems can support health professionals' decision-making by offering guidelines or forcing steps to minimize human errors (Aung et al., 2021). During surgical procedures, AI can guide robotic equipment to improve the precision of the human movement (Davenport TH, Glaser, 2002; Hashimoto et al., 2020) or analyze data from the operating field, providing advice and clean visualizations in real-time (Shaheen MY, 2021). Telemedicine applications can support treatment programs with remote reminders, data collection, and analysis (Azzi et al., 2020) (Kitsios et al., 2023).

Aside from clinical practice, healthcare facilities can adopt AI solutions for cybersecurity, management, research, and training. Cybersecurity automation aims to keep the digital system and data safe. In management and administration, AI-based process analysis can find weaknesses and bottlenecks in the organization and provide operational advice. AI software can support supply acquisition, revenue cycles, and visit scheduling (Bhakoo et al., 2012). Machine learning solutions can support data analysis during clinical trials (Al Kuwaiti et al., 2023). AI-based virtual and mixed reality simulations can be used to train medical students (Sharma & Jindal, 2023) (Secinaro et al., 2021).

### 2.2 Stakeholders

Healthcare is a highly complex ecosystem aiming at promoting, improving, and re-establishing the health conditions of a specific population. This ecosystem encompasses different types of stakeholders, i.e., people and organizations who influence or are influenced by the process at stake (Freeman, 2011). The information gained from a stakeholder analysis is valuable when seeking a collaborative approach to planning, developing, and delivering healthcare services and innovations. It allows planners to understand the different perspectives on such innovation (Guise et al., 2024).

Various studies have sought to map healthcare stakeholders (An et al., 2022) (Austen & Frąckiewicz, 2018) (Boonstra & Govers, 2009) (Frączkiewicz-Wronka et al., 2021); (Guise et al., 2024); (Moloi et al., 2022); (Nieder et al., 2020); (Tampio et al., 2022). The analysis of these studies identifies several classes of stakeholders who, following a simplified model by Kumar et al. (Kumar et al., 2023), can accommodate three concentric circles. In the outermost circle are stakeholders building the technical, administrative, and regulatory infrastructure for the use of AI. They include local and national ministries, policymakers, funding bodies and donors, health education agencies, professional orders, trade unions, and competing health facilities. In an inner circle are stakeholders who, although external to the health facility, interact directly with it; they include patients' families, associations and organizations for families and/or patients, counselling services, cryogenic banks, physicians' and psychotherapists' private practice, pharmaceutical companies, suppliers, and insurance companies.

The innermost circle consists of four groups of stakeholders within the healthcare facility: (a) patients who are hospitalized, use telemedicine devices from home, or participate in clinical trials; (b) health professionals including doctors, nurses, paramedical staff, and X-ray technicians; (c) hospital administrators and executives; (d) non-clinical hospital units such as technical department, communication with the public and advertising. These four classes of direct stakeholders will be included in our study.

### 3. Transparency

#### 3.1 Legal transparency requirements

International regulatory and standardization bodies unanimously advocate transparency as one of the building blocks of a trustworthy AI (Cousineau et al., 2025). In the EU regulation for trustworthy AI, or the AI Act, transparency is broken down into traceability, explainability, and communication (Masotina et al., 2023). Traceability consists of documenting the datasets and processes at the basis of AI output; explainability refers to providing information about the criteria and mechanism leading to its output; and communication consists of signalling to users the presence of an AI and its capabilities. Transparency also provides a necessary condition for applying another ethical principle, i.e., human control over AI, and exercising a right, i.e., the contestation of unjustly unfavorable output (Almada, 2019). Unlike other ethical principles for trustworthy AI, transparency empowers the users.

In the case of high-risk devices, whose malfunctioning can directly damage the users' rights and well-being, the legislator imposes stringent and non-discretionary transparency obligations. Among them is the preparation of a notice, or IFU, which must include 12 prescribed types of information and address the deployers (article 13 of the EU AI Act). Moreover, the regulation imposes that such information be understandable: "Providers should ensure that all documentation, including the instructions for use, contains meaningful, comprehensive, accessible and understandable information, taking into account the needs and foreseeable knowledge of the target deployers (Artificial Intelligence Act, 2024)." In a sociotechnical approach to technology (Button, 1993) (Gagliardi, 1990), the characteristics of a computer artifact depend not only on its technical functionalities but also on the user's objectives. Thus, explaining such artifacts requires reaching out to the intended deployers and considering what aspects are relevant for them, given their roles and usage context. A space is therefore created to investigate the deployers' transparency needs and fine-tune the IFU to those needs.

#### 3.2 User-based transparency needs

While legal transparency explains the AI systems' processes and sources (Singhal et al., 2024), user-based transparency allows users to use the AI system confidently. Although the goal is not to persuade them to accept an AI, the studies of the users' reservations about adopting AI in healthcare can provide insights into the kind of honest information they might need.

In a survey (Al Harbi et al., 2024) with 109 anaesthesiologists in Saudi Arabia, most respondents agreed that the AI introduction should be gradual (82.57%). When imagining a primary care consultation involving both doctor and AI (Kocaballi A.B. et al., 2020), 16 primary care doctors highlighted the importance that clinical reasoning remained their core task, along with conversation and empathy. They also wished the AI assistant would adapt to the doctors' working style and consultation workflow, including generating patient summary letters. Witkowski and colleagues surveyed 600 US-based patients representative of the Florida population in terms of age, gender, race/ethnicity, and political affiliation (Witkowski et al., 2024). Among the sample's concerns was the fear of losing the human element of the medical care; thus, the respondents were comfortable with AI deployed to schedule appointments (84,2%), less with AI collecting their medical history (60,7%) or predicting future medical conditions (52%). Sangers (Sangers et al., 2021) investigated the perception of mobile health applications incorporating artificial Intelligence for skin cancer screening with 27 patients; the elements facilitating or hampering their use included the developer's trustworthiness, the endorsement of the healthcare provider, and their cost.

A synthesis of the users' concerns is provided by Vo and colleagues (Vo et al., 2023). They systematically reviewed the scientific literature on AI in healthcare, looking for qualitative and quantitative studies with health professionals, patients, and the general population. They selected and analyzed 105 studies from 18 countries published between 2017 and 2021 and extracted 271 themes, further grouped into seven categories: (i) knowledge and familiarity of AI, (ii) AI benefits, risks, and challenges, (iii) AI acceptability, (iv) AI development, (v) AI implementation, (vi) AI regulations, and (vii) Human – AI relationship. By browsing through the long list of themes provided by the authors, some themes repeat multiple times. For instance, understanding AI can be found in at least three categories: as an element promoting the acceptability of AI, as a perceived requirement during AI development, and as a possible AI risk. Themes also recur within a single category; for example, the fact that human health experts need to keep relating to patients is listed as an AI-related factor increasing acceptance, a motivation for negative attitudes, and a motivation for supportive attitudes. This redundancy depends on the different assumptions of the studies reviewed, where the same construct is connected to different variables. It also depends on the fact that the positive or negative versions of the same concern were treated as distinct factors: for instance, "Replacing human doctors completely or partially" was considered different from its opposite, "not replacing human doctors." In their discussion, however, the authors boil down all themes to a smaller set: data privacy, preservation of clinical judgment, attention towards possible health disparities, empathic and personalized care, accuracy, need for education and training, clear legal responsibilities, human agency, presence of regulatory standards, validation before large-scale deployment, fairness, accountability, transparency, and ethics. This more concise set will inspire our study's list of transparency needs.

## 4. Study

This study tests how stakeholders of healthcare AI value and map users' transparency needs into the IFU structure. By stakeholders, we mean the four types of direct stakeholders identified in section 2.2; by transparency needs, we mean those highlighted in previous studies and described in section 3.2. The part of the IFU we evaluate is the table of contents, which consists of 12 mandatory types of information foreseen by Article 13 of the AI Act. Our research questions, thus, read as follows:

RQ1: When presented with a transparency need, which IFU sections do participants identify as the source of information addressing that need?

RQ2: How relevant are the transparency needs to the four types of direct stakeholders?

### 4.1 Material

#### 4.1.1 Simplified IFU

Starting from the description of the IFU sections contained in the AI Act, we built a simplified version suiting an online survey. We removed articles, repeated references to provider, deployer, or AI systems, adverbs expressing conditions (i.e., where relevant, if any, where applicable), non-informative modifiers or heads of modifiers (a description of…), and references to other articles in the Act. We also streamlined the structure with pre-modifications (e.g., "needed hardware resources" instead of "hardware resources needed"), and itemizations (e.g., bullet points).

We then assigned each section an icon. To our knowledge, there is no set of validated icons for AI transparency compliance, apart from work by Friederich and colleagues, which is very specific to AI data (Lancaster University, United Kingdom et al., 2020). So we relied on privacy icons (Rossi & Palmirani, 2018), finding only one applicable to IFU sections, i.e., "intended purpose." For all other sections, we relied on the MS Office icon library, looking for icons that were semantically compatible with the IFU section titles.

The sections were finally grouped into two blocks, singling out those related to the AI system's capabilities and limitations according to Article 13. The resulting IFU structure is illustrated in Figure 1. Its relationship with the original wording in the AI Act is reported in the Appendix (Table 4).

#### 4.1.2 Questionnaire

The transparency needs used in the study are adapted from (Vo et al., 2023) and are listed in Table 1. The needs are grouped into four broad and intuitive categories: accuracy, humanization, law, and resources.

**Table 1.** The list of user-based transparency needs related to medical AI (adapted from (Vo et al., 2023).

| Type | Abbrev. | Content |
| --- | --- | --- |
| Accuracy | a - bs | The AI device might not have been tested on all patient types. What patient types are underrepresented? |
| Accuracy | a - cl | Each medical situation is complex. What characteristics does the AI take into account of each medical situation? |
| Accuracy | a - cm | The AI might not suit all patients. Are there patients for which this device is not recommended? |
| Accuracy | a - pp | The AI device needs to perform its intended job reliably. Was it tested to ensure it meets expectations? |
| Accuracy | a - pr pt | This AI device must have been in use long enough. On how many patients has it been used? |
| Accuracy | a - ps | I wonder what patients will feel. Were past patients happy with the care provided with the AI device? |
| Humanization | h - br | Doctors have many tasks besides treatment, such as creating summary letters for patients. Does the AI device have features that assist in all aspects of their job? |
| Humanization | h - cn | Each patient is unique. Does the AI device consider the unique characteristics of each patient? |
| Humanization | h - ov | The AI device might malfunction or make mistakes. Is there a person overseeing the AI device? |
| Humanization | h - pr | Language, culture, or financial barriers could make using the AI device challenging. What are the requirements to use it? |
| Humanization | h - sp | I might not understand the AI output; does the device come with simple explanation of what its results mean? |
| Humanization | h - un | Human contact is important in clinical practice. Do doctors and patients still interact with each other when this AI device is used? |
| Resources | l - adv | Treatments with this AI device seem unusual and fancy. Do insurance companies cover treatments that use it? |
| Resources | l - cy | The AI device should work with the healthcare system in my region. Is the AI device connected to it? |
| Resources | l - dt | The AI equipment and research are expensive. Is there any funding to cover the high costs? |
| Resources | l - l | I do not want to be left alone to figure out how to use the AI device. Is there any training available to help users operate it? |
| Resources | l - lb | Ease of use is essential for any device. Was the system tested to ensure it is user-friendly? |
| Resources | l - ms | The proper functioning of hardware and software, a stable connection, and access to technical support are crucial for the device's reliability. What guarantees are there for these aspects? |
| Law | r - fn | The AI device could be misused to alter or take advantage of health results. Are there measures in place to prevent this? |
| Law | r - fu | There is always a chance that something can go wrong with a treatment. If something happens, is the responsibility for the mistake with the AI device or the user? |
| Law | r - hl | Unauthorized outsiders might try to access the device; is it protected against that? |
| Law | r - in | The AI device surely collects my personal data and health information. Does this AI device keep it protected? |
| Law | r - tr | Some digital devices include advertisements. Does this one have any? |
| Law | r - us | For the device to be legal, it must follow certain regulations and official guidelines. What are they? |

In the questionnaire, we only displayed two transparency needs to avoid careless answers from fatigued participants. The following information was collected for each transparency need: perceived relevance, localizability, and confidence level. More specifically:
- Perceived relevance was measured with the item "How relevant is that information to trustfully use an AI device?" on a 5-degree scale from 5 (absolutely relevant) to 1 (totally irrelevant)."
- Localizability was collected with a multiple-choice question asking, "Select the section most likely to contain an answer to the user's query," followed by the 12 IFU (Figure 1). An open-ended question asking the participant to motivate the selection ("Please explain the reasons for selecting that IFU section") was included. It was meant to encourage a pondered selection and was not analyzed.
- Confidence was measured with the item "How confident are you that it is the right section?" on a 5-degree scale from 5 (very confident) to 1 (very doubtful).

One additional item measured the participants' familiarity with the lexicon used in the IFU structure. The item displayed the titles of the 12 IFU sections; the participants were asked to highlight any obscure words. They had no restrictions on the number of words, including highlighting none (in this case, a confirmation message appeared to ensure the lack of selection was deliberate). All words could be marked except articles, conjunctions, and prepositions.

### 4.2 Setting

The questionnaire was built and administered via the online survey platform Qualtrics (https://Qualtrics.com). No restrictions were placed on the device to be used, and the survey format was optimized for mobile devices according to Qualtrics guidelines (e.g., using a multiple-choice question format instead of matrices, no more than two answer columns, and the vertical alignment of the answer options). The questionnaire included a few strategies to maximize the quality of the data collected. The instruction pages and question blocks were timed to discourage rushed answers so that participants could not proceed in the questionnaire until a few seconds had passed. An item prevented the participants who ignored the introductory part of the survey from continuing. It consisted of a multiple-choice question asking, "So, what is an IFU?" with three answer options: "A document with instructions for the users of an AI device" (correct), "An AI device supporting physicians in healthcare" or "The name of a recent EU regulation regarding healthcare." Only data from participants recognizing the correct answer were analyzed. The deactivation of multiple submissions from the same ID and the inclusion of a bot check ("Captcha") prevented respondents from using software that automatically fills in forms (Xu et al., 2022). Steps were also taken to prevent open questions from being answered using LLMs: LLMs were explicitly forbidden in the instructions and technically made difficult by removing the ability to copy and paste text from/to the survey. This combination discourages LLMs' usage better than other strategies (Veselovsky et al., 2025).

### 4.3 Procedure

The protocol started with the invitation sent by the recruitment platform to all eligible subscribers: "We are investigating the information needed to use an AI system trustfully. If you participate, you will rate the relevance of such information and locate it in the table of contents of the instructions sheet. We are very interested in your answers, so the survey includes attention checks and adopts technical deterrents against using LLM to write answers. Completing the survey and passing attention checks is compensated with 3 UK£." The invitees interested in participating were directed to the short version of the information note. From there, they could download the complete, detailed note, if they wished. Consent was expressed by pressing the "I agree" option at the bottom of the information note.

The participants who agreed to participate were then briefly introduced to the topic of the study: "Artificial Intelligence technologies can support healthcare professionals in tasks such as diagnosis, information retrieval, treatment decisions, surgical operations, and the analysis of large amounts of data during trials. A mandatory document called 'IFU' ('Instructions for Use') accompanies any AI device and addresses its users. Users can be hospital administrators, physicians, technicians, or patients. An IFU contains 12 sections."

The IFU structure was then displayed, along with the request to highlight any obscure word. The attention check was then administered.

**Fig 1** An instance of a block with a transparency need and four items collecting relevance, ease to find, motivation, and confidence

The participants were then given the following instructions: "We will show 2 questions the AI user might have. 1. Read the question. 2. Judge its relevance. 3. Select the IFU section possibly containing the answer. It takes 30 seconds before the second question is displayed to prevent rushed answers. Thank you in advance for your time on this project!" After 10 seconds since the instructions were first shown, a forward button became available, and the participants were presented with a first block of questions related to a transparency need. The block was selected by two subsequent automated randomizers, one choosing the type of transparency need (i.e., accuracy, humanization, law, and resources) and a second choosing a transparency need among the six available in each type. The randomizers were set to display all needs the same amount of time across each subsample.

A second block was displayed when the participant completed the first block and at least 30 seconds had passed. A timer was included, which was set to count up instead of down to avoid suggesting a time limit.

### 4.4 Pilots

We ran the survey with 12 participants and added an item asking for feedback to improve the survey. These pilots suggested some adjustments to the way the survey mechanism was initially set. We applied a concern-and-question format to the transparency needs' description, to better highlight the concern they stem from (e.g., "The AI device could be misused to alter or take advantage of health results. Are there measures in place to prevent this?" instead of "Are there measures to prevent the AI device from being used to alter or take advantage of health results?"). Moreover, we randomized complete blocks instead of randomizing the transparency needs only, to easily refer the scores to specific needs. We also autoloaded instructions within the same page instead of automatically progressing through instruction pages to accommodate every reading speed.

### 4.5 Ethics

The study was approved by the Ethical Committee of the Human Inspired Technologies Research Center (HIT) (2024_263R1).

### 4.6 Data analysis

The data allowed to obtain the following measures:
- Number of obscure terms per participant.
- Overall frequency with which each term was highlighted.
- Selection frequency of an IFU section per transparency need.
- Confidence scores per selection.
- Relevance score per transparency need.

In addition, the recruitment platform provided information about the participant's biological sex, age, country of residence, and occupation.

The results are summarized with descriptive statistics.

### 4.7 Participants

The recruitment was carried out via the online platform Prolific (https://www.prolific.com). Participants were adults residing in the European Economic Area, where the AI Act is enforced. Balanced quotas were recruited by biological sex, with fluent knowledge of English. Pilot participants were excluded from the main study. Further inclusion criteria created four subsamples: (a) doctors, nurses, and paramedics; (b) managers and directors; (c) workers in the IT sector; and (d) patients (people with chronic illness or disease and no medical education). A total of 1079 respondents took the survey. Incomplete surveys by respondents who did not pass the attention check, experienced some technical issue, or withdrew were removed. The final sample comprised 813 respondents, whose sociodemographic characteristics are shown in Table 2.

**Table 2**. Demographic characteristics of the sample

| | N | age Mean (SD) | biol. sex f | biol. sex m | job | country of residence |
|---|---|---|---|---|---|---|
| managers | 238 (29.31%) | 35.7 (10.1) | 117 | 121 | Director (7.56%), Manager (92.44%) | 20 (Germany 11.76%, Italy 12.18%, Poland 16.81%, Portugal 17.65%, other 41,6%) |
| health prof. | 115 (14.16%) | 31.2 (9.3) | 61 | 54 | Doctor (62.61%), Nurse (35.65%), paramedic (1.74%) | 16 (Germany 13.04%, Italy 10.43%, Poland 13.91%, Portugal 21.74%, Other 40.88%) |
| patients | 229 (28.2%) | 25.9 (5.1) | 114 | 115 | - | 20 (Poland 31.44%, Portugal 15.72%, other 52.84%) |
| tech. | 230 (28.33%) | 33.0 (9.1) | 107 | 123 | IT (53.04%), other (46.96%) | 20 (Germany 13.91%, Italy 12.17%, Poland 16.52%, Portugal 15.65%, other 41,75%) |
| tot | 812 | 31.5 (9.4) | 399 | 413 | | |

The gender distribution was balanced in the subsamples. Half of the sample resided in Germany, Italy, Poland, or Portugal, and the rest in 15 other countries in the European Economic Area. The managers' subsample was mainly composed of people in the roles of manager and accountant, the health professionals subsample of doctors and nurses, and the tech subsample distributed across various information technology roles (Appendix, Table 5).

Regarding the presence of obscure words in our simplified IFU structure, 85% of the respondents highlighted none to two words. The participants highlighting obscure words were well distributed across subsamples (admin 29%, health professionals 14%, patients 28%, and tech 28%). The words mentioned more frequently were "resilience" ($n = 203$) and "computational" ($n = 144$). The complete list is in the Appendix (Table 6).

## 5. Results

### 5.1 Need-section mapping

Table 3 reports the selection frequency of the different IFU sections per transparency need. In each line, the highest percentage indicates the section that was more often selected with reference to a transparency need. By reverse engineering these results, IFU writers would know where information can be placed to maximize localizability. For example, our percentages suggest the following allocations:

- In Section 1 "Intended purpose of the AI system," details on how the AI system supports the health professional's practical tasks (h – pr).
- In Section 2 "Correctness, performance resilience, and cybersecurity," the evidence about the AI system's reliability (a – cl) and its connection to the regional healthcare system (r – hl, although this last point can be included in sections 8 or 9). This section could also mention which measures it incorporates to prevent intrusions from unauthorized outsiders (l – cy) and misuse or fraud (l - l). Here could also be cited the regulations and guidelines it must comply with (l – l).
- In Section 3, "Risks to health, safety, and fundamental rights," the legal responsibilities in case a mistake is made using the device (l – lb) and the insurance coverage for medical procedures supported by the AI system (r – in).
- In Section 4, "Capabilities to explain its results," how the AI system explains its output (h – un).
- In Section 5, "Performance with specific (groups of) persons," the types of patients with whom the system was trained or validated (a – bs) and are good candidates for treatment with the AI (a – pp); the previous patients' satisfaction with the AI system (a – ps); any linguistic, cultural and financial requirements to use the system (h – br), and how each patient's uniqueness is taken into account when relying on the AI system (h – sp).
- In Section 6, "Datasets used to train, validate and test the system," how the AI system handles the complexity of each medical situation without oversimplifying it (a – cm) and the number of patients with which it has already been used (a – pr pt).
- In Section 7, "Proper usage," existing training for end users (r – tr) and document the AI system's usability (r – us)."
- In Section 8, "Provider's identity, contact details, and authorized representative," like in Section 2, how the AI system is connected with the healthcare system in use in the region (r – hl).
- In Section 9, "Collection, storage, and interpretation of activities' record," alert about the presence of advertisements (l -adv) and describe the provisions to protect personal data (l – dt).
- In Section 10, "Needed computational and hardware resources, maintenance/care and expected lifetime," the guarantees to proper functioning and technical support (r – fn) and the source of the funding keeping the system running (r – fu).
- Section 11, "Planned changes to the AI system," was the only section associated with none of the transparency needs we proposed to participants.
- In Section 12, "Human control over the machine, including support in interpreting the results," how human contact with patients is preserved while using the AI system (h – cn) and human oversight is implemented (h – ov).

The same data also suggests some possible issues, as illustrated by the shape of the frequency distribution in Figure 2. A few lines have one clear peak; several diagrams, however, have multiple or no peaks. When peaks are low or multiple, the preferred IFU section barely reaches 50% of the sample. This occurs in 16 cases (a – cl, a – cm, a – pp, a pr pt, h – br, h – cn, h – pr, l – adv, l – l, l – ms, r – hl, r – in, r – tr, r – us), equaling 66% of the needs under evaluation. Curiously, the respondents did not report any sense of uncertainty even in the most equivocal cases: for example, the two needs marked as "r – in" and "l – adv" are among the transparency needs more largely distributed across IFU sections, but respondents declared a confidence level above 4. This pervasive, undetected equivocality suggests that a user-centered IFU should be integrated or prefaced with a section working as an interface between the users and the content, directing them to the IFU section answering their questions.

**Table 3.** Selection frequency (in percentage) of each IFU section per transparency. The highest value in each row is boldfaced (framed if < 50). Perceived confidence from 1 to 5, where 5 = very confident.

| | Selection frequency (%) | | | | | | | | | | | | N | Perceived confidence | |
|---|---|---|---|---|---|---|---|---|---|---|---|---|---|---|---|
| | IFU sections | | | | | | | | | | | | | | |
| Transp. needs | 1 | 2 | 3 | 4 | 5 | 6 | 7 | 8 | 9 | 10 | 11 | 12 | | M | SD |
| a - bs | 0 | 0 | 6 | 4 | **57** | 19 | 0 | 1 | 7 | 1 | 1 | 3 | 70 | 3.72 | 0.81 |
| a - cl | 7 | 47 | 0 | 4 | 3 | 28 | 3 | 0 | 1 | 0 | 1 | 4 | 68 | 4.13 | 0.86 |
| a - cm | 6 | 18 | 10 | 12 | 6 | 19 | 4 | 3 | 7 | 1 | 0 | 12 | 67 | 2.89 | 1.08 |
| a - pp | 13 | 2 | 21 | 3 | 44 | 0 | 2 | 3 | 5 | 0 | 2 | 6 | 63 | 3.73 | 0.82 |
| a - pr pt | 0 | 3 | 4 | 0 | 16 | 42 | 3 | 0 | 27 | 3 | 0 | 1 | 67 | 3.30 | 0.93 |
| a - ps | 1 | 10 | 6 | 6 | **54** | 0 | 3 | 0 | 10 | 0 | 0 | 9 | 67 | 3.49 | 0.94 |
| h - br | 4 | 4 | 4 | 3 | 38 | 4 | 22 | 3 | 0 | 6 | 1 | 9 | 68 | 3.31 | 0.97 |
| h - cn | 16 | 0 | 6 | 6 | 1 | 0 | 24 | 0 | 3 | 0 | 0 | 44 | 68 | 3.19 | 0.97 |
| h - ov | 1 | 9 | 3 | 3 | 0 | 0 | 1 | 4 | 1 | 4 | 1 | **71** | 69 | 3.61 | 0.80 |
| h - pr | 47 | 3 | 0 | 22 | 0 | 1 | 6 | 3 | 9 | 1 | 0 | 7 | 68 | 3.15 | 1.00 |
| h - sp | 3 | 10 | 1 | 3 | **51** | 7 | 4 | 7 | 3 | 0 | 1 | 7 | 67 | 3.57 | 0.95 |
| h - un | 1 | 3 | 0 | **63** | 0 | 3 | 3 | 1 | 1 | 1 | 1 | 21 | 68 | 3.75 | 0.92 |
| l - adv | 16 | 10 | 0 | 0 | 6 | 3 | 11 | 21 | 24 | 3 | 2 | 5 | 63 | 4.07 | 0.77 |
| l - cy | 0 | **71** | 6 | 0 | 0 | 0 | 1 | 12 | 4 | 0 | 0 | 6 | 69 | 3.91 | 0.85 |
| l - dt | 0 | 25 | 7 | 0 | 0 | 0 | 3 | 4 | **61** | 0 | 0 | 0 | 69 | 4.09 | 0.73 |
| l - l | 11 | 32 | 9 | 0 | 0 | 5 | 15 | 17 | 0 | 3 | 2 | 8 | 66 | 3.88 | 0.73 |
| l - lb | 0 | 4 | **50** | 0 | 0 | 0 | 4 | 1 | 1 | 0 | 0 | 38 | 68 | 3.51 | 0.88 |
| l - ms | 4 | 30 | 15 | 3 | 1 | 3 | 18 | 0 | 7 | 0 | 1 | 16 | 67 | 3.78 | 0.96 |
| r - fn | 1 | 19 | 0 | 0 | 0 | 1 | 7 | 3 | 1 | **57** | 0 | 9 | 68 | 3.96 | 0.80 |
| r - fu | 3 | 5 | 0 | 2 | 0 | 0 | 3 | 27 | 0 | **58** | 2 | 2 | 66 | 3.55 | 0.96 |
| r - hl | 0 | 15 | 1 | 4 | 6 | 7 | 6 | 15 | 15 | 12 | 0 | 1 | 68 | 3.36 | 1.03 |
| r - in | 23 | 3 | 29 | 5 | 2 | 0 | 8 | 12 | 0 | 5 | 5 | 11 | 66 | 4.38 | 0.69 |
| r - tr | 4 | 4 | 1 | 0 | 1 | 6 | 46 | 0 | 0 | 1 | 1 | 34 | 70 | 4.28 | 0.78 |
| r - us | 3 | 7 | 3 | 7 | 14 | 10 | 36 | 0 | 1 | 3 | 1 | 14 | 70 | 3.84 | 0.77 |

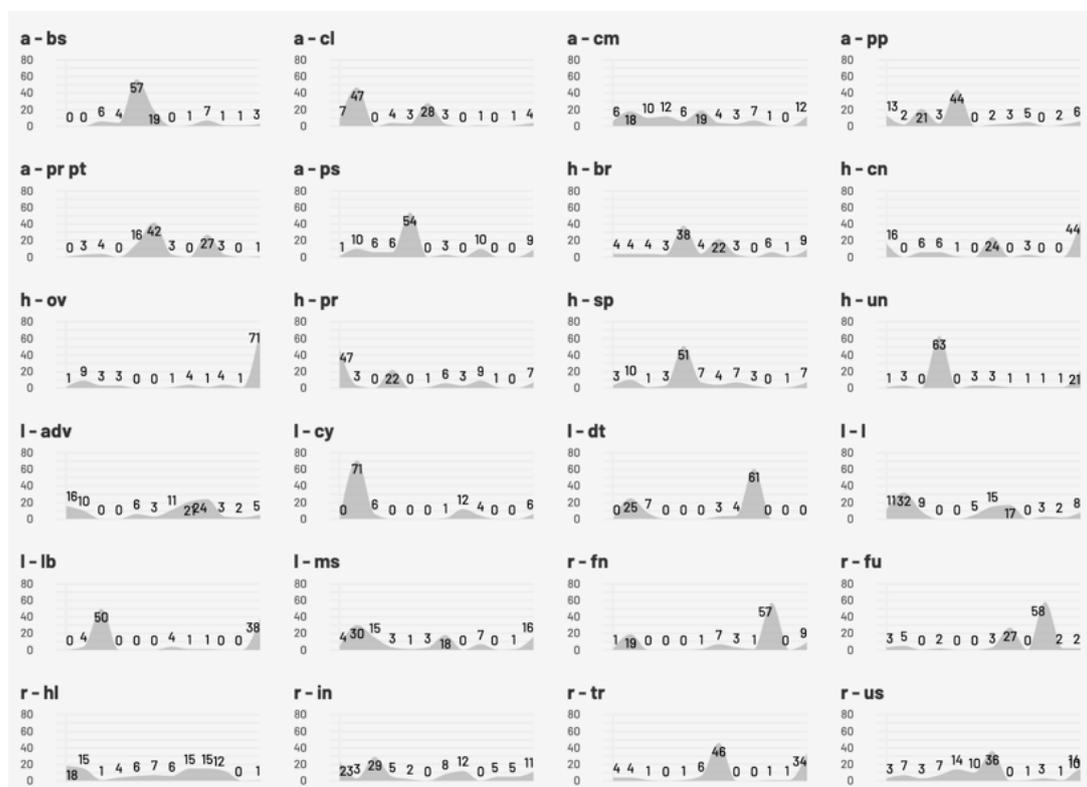

**Fig 2** The diagrams illustrate the frequency with which each section was chosen as the probable repository of the information for each transparency need. The data are the same as in Table 3 (these diagrams were created with https://flourish.studio/)

## 5.2 Need's relevance

Figure 3 reports the relevance scores attributed to the transparency needs. All values are higher than the mid value of the scale (2.5), suggesting that all transparency needs were considered relevant. The choice of needs included in this survey is therefore validated. A visual inspection also suggests that a few transparency needs remain relevant across stakeholder types. They include the information about data protection (l – dt), liability in case of mistakes (l – lb), previous patients' satisfaction (a – ps), and, at the bottom, funds (r – fu) and the presence of advertising (l - adv). Instead, other needs' relevance scores vary across stakeholders, creating steep slopes in the lines connecting the scores. For example, knowing which types of patients are underrepresented in the AI (a – bs) is a top transparency need for health professionals and patients, but is 9th and 15th for managers and technical stakeholders. The AI device's usability (r - us) and the availability of training (r – tr) seem highly relevant to health professionals, but less so to technicians and managers. The connection with the regional health system (r - hl) and insurance companies (r – in) seems important to health professionals and technicians, but less so for other users. Thus, each column shows the priorities of a specific stakeholder. Moving from managers to health professionals, the ability to meet expectations (a-cl), the AI proper functioning (r – fn), the reference to relevant guidelines and regulations (l – l) decline in relevance, while patients' underrepresentation (a – bs), device usability (r - us), connection with the regional health system (r - hl) and insurance companies (r – in) increase in relevance. These scores suggest that different deployers might need some information to be highlighted for them, because it is more relevant to their role and usage goals.

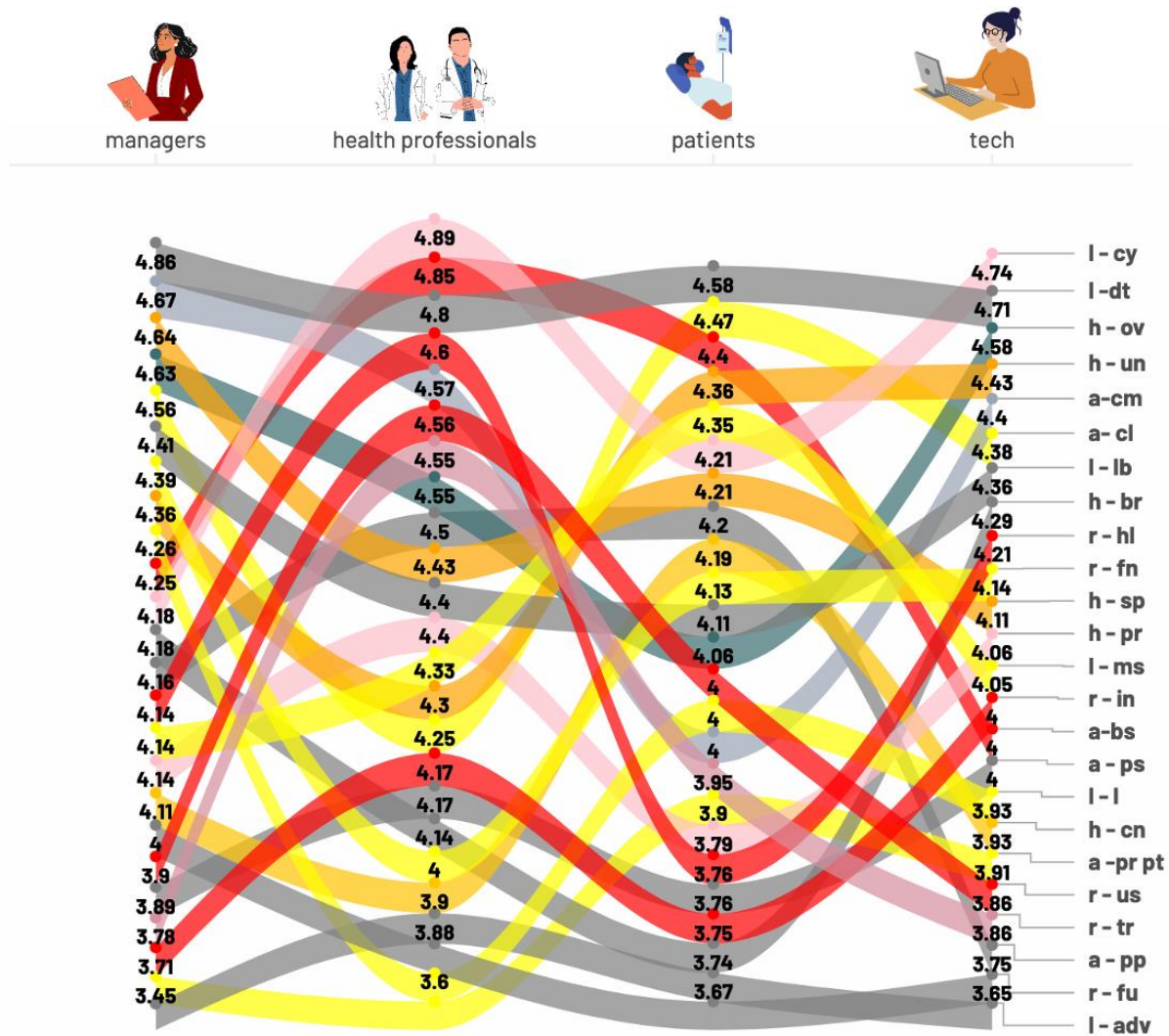

**Fig 3** The transparency needs' rankings (average) in each subsample. Red/pink lines have a high peak with health professionals; yellow/orange lines have a high peak with managers and patients. Grey and green have flat or peculiar peaks. (This diagram was created with https://flourish.studio/)

## 6. Discussion

Our study suggests that the transparency needs map onto the IFU structure in different ways. Some of them map directly onto a specific IFU section; they seem isomorphic in terms of topic and level of abstraction. For example, the protection against unauthorized users maps very well onto the section "Correctness, performance resilience and cybersecurity;" likewise, information on human oversight maps well onto the section "Human control over the machine." In both cases, the sections were chosen by 71% of the participants. Other needs have no isomorphism with a specific IFU section and could fit many. For instance, patients for whom the AI-supported treatment is inadequate could be equally described under "Intended purpose," "Risks to health," or "Performance with specific groups of persons." Likewise, whether there are previous clinical trials on the medical device could be categorized under "Correctness" as much as "Datasets used to train, validate, and test the system". Finally, there are information needs that do not seem to be map onto any IFU section, e.g., whether the liability lies with the machine in the event of an error, what are the relevant regulations for the use of the device, whether human contact is preserved or whether the complexity of the medical case is respected. Although these are recurrent transparency needs in healthcare AI, the legal IFU does not seem equipped to satisfy them.

The way the IFU is modelled by the regulators should not be considered as the finalized interface of the document presented to the users. Based on the test results reported in this paper, we suggest integrating the mandatory information within a more encompassing and isomorphic document to the users' information needs. This general recommendation breaks down into five strategies:

1. *Adding navigation interfaces*. Gils and colleagues, based on their legal design workshops (Gils et al., 2024), Gils and colleagues recommend using a logical structure and table of contents. The IFU needs an additional layer to access its information according to the users' transparency needs.

2. *Diversifying access by stakeholders' priorities.* IFU information can also be navigated differently based on the type of stakeholder who approaches it with their information priorities. Investigating different stakeholders reveals such priorities and can help design the information interface accordingly.

3. *Locally adding case-specific information*. Several transparency needs are specific to the context in which the AI system is deployed. The provider can then include signposts for this information to be filled in locally.

4. *Offering reference to other documents and sources*. An IFU cannot exhaust all information needs, nor does the documentation need to be bulky and redundant; thus, the IFU can include pointers to information handled elsewhere, e.g., regulations. Gils et al (Gils et al., 2024) have a similar remark about pointers to technical manuals.

5. *Empirically checking information localizability.* User studies can help prepare a usable IFU and support the drafting process by identifying information needs and the section most suitable to host the related information. They can also guide the assessment of a finalized draft and measure the ease with which stakeholders can locate needed information.

In addition to these specific recommendations, the IFU can follow design guidelines to make the document more readable and comprehensible, as discussed in existing literature and practical guides. For example, Crisan et al. (Crisan et al., 2022) suggest that notices and cards should include warnings, prompts, and summaries, which clarify how to act on the information provided. Seifert and colleagues (Seifert et al., 2019) propose labels conveying the transparency level of an AI system.

### 6.1 Limits

This study has several limits. It does not describe what the participants understood of the content of the sections. This would have required a different setup with comprehension checks. Also, we did not investigate why certain words were considered obscure. Managers and techs were not employed explicitly in hospital facilities. Finally, the information needs in our study were very generic, as were the subsamples, to prove a concept; IFUs for specific AI technologies will need to be tested with more specific end users.

### Conclusions

It is of general societal interest that IFUs achieve more than formal compliance with a regulatory prescription. The effectiveness of such a document should then be measured in terms of the presence of relevant information and the ease with which its readers retrieve it. To facilitate this goal, we have attempted to map some of the most frequently reported transparency needs into the IFU sections. The results suggest that IFU needs local integration in terms of design and type of information before it becomes fully useful to stakeholders. This approach can easily be replicated and exported to other high-risk systems to help achieve the sort of meaningful transparency wished for by the regulators.

**Acknowledgments**

The authors would like to thank Nicolò Navarin for his comments about medical AI. This study was partially supported by the European Union NextGenerationEU program through the project "Beyond compliance: AI act made usable in Healthcare - UseAI" (Project n.: J33C22002830006, Extended Partnership Future Artificial Intelligence Research – FAIR, Spoke 8 Pervasive AI).

**CRediT authorship contribution statement**

Anna Spagnolli: Conceptualization, Methodology, Investigation, Data curation, Formal analysis, Funding acquisition, Writing - Review & Editing. Cecilia Tolomini: Writing – original draft, Resources. Elisa Beretta: Resources. Claudio Sarra: Conceptualization, Funding acquisition.

**Data availability statement**

The anonymous, quantitative dataset generated by the survey during the current study is available in Zenodo public repository (https://doi.org/10.5281/zenodo.15476368).

**Statements and declarations**

The authors declare no competing interests.

Chellasamy, A., & Nagarathinam, A. (2022). An Overview of Augmenting AI Application in Healthcare. In A. P. Pandian, X. Fernando, & W. Haoxiang (Eds.), *Computer Networks, Big Data and IoT* (Vol. 117, pp. 397–407). Springer Nature Singapore. https://doi.org/10.1007/978-981-19-0898-9_31

Chustecki, M. (2024). Benefits and Risks of AI in Health Care: Narrative Review. *Interactive Journal of Medical Research*, *13*, e53616. https://doi.org/10.2196/53616

Cousineau, C., Herger, N., & Dara, R. (2025). Transparency requirements across AI legislative acts, frameworks and organizations: Shaping a sample transparency card. *AI and Ethics*. https://doi.org/10.1007/s43681-025-00725-5

Crisan, A., Drouhard, M., Vig, J., & Rajani, N. (2022). Interactive Model Cards: A Human-Centered Approach to Model Documentation. *2022 ACM Conference on Fairness, Accountability, and Transparency*, 427–439. https://doi.org/10.1145/3531146.3533108

Frączkiewicz-Wronka, A., Ingram, T., Szymaniec-Mlicka, K., & Tworek, P. (2021). Risk Management and Financial Stability in the Polish Public Hospitals: The Moderating Effect of the Stakeholders' Engagement in the Decision-Making. *Risks*, *9*(5), 87. https://doi.org/10.3390/risks9050087

Freeman, R. E. (2011). *Strategic management: A stakeholder approach* (Nachdr.). Cambridge Univ. Press.

Gagliardi, P. (Ed.). (1990). *Symbols and Artifacts: Views of the Corporate Landscape*. De Gruyter.

Gils, T., Heymans, F., Ooms, W., & De Bruyne, J. (2024). From Policy to Practice: Prototyping The EU AI Act's Transparency Requirements. *SSRN Electronic Journal*. https://doi.org/10.2139/ssrn.4714345

Guise, V., Chambers, M., Lyng, H. B., Haraldseid-Driftland, C., Schibevaag, L., Fagerdal, B., Dombestein, H., Ree, E., & Wiig, S. (2024). Identifying, categorising, and mapping actors involved in resilience in healthcare: A qualitative stakeholder analysis. *BMC Health Services Research*, *24*(1), 230. https://doi.org/10.1186/s12913-024-10654-4

Kitsios, F., Kamariotou, M., Syngelakis, A. I., & Talias, M. A. (2023). Recent Advances of Artificial Intelligence in Healthcare: A Systematic Literature Review. *Applied Sciences*, *13*(13), 7479. https://doi.org/10.3390/app13137479

Kocaballi A.B., A., Ijaz, K., Laranjo, L., Quiroz, J. C., Rezazadegan, D., Tong, H. L., Willcock, S., Berkovsky, S., & Coiera, E. (2020). Envisioning an artificial intelligence documentation assistant for future primary care consultations: A co-design study with general practitioners. *Journal of the American Medical Informatics Association*, *27*(11), 1695–1704. https://doi.org/10.1093/jamia/ocaa131

Kumar, P., Chauhan, S., & Awasthi, L. K. (2023). Artificial Intelligence in Healthcare: Review, Ethics, Trust Challenges & Future Research Directions. *Engineering Applications of Artificial Intelligence*, *120*, 105894. https://doi.org/10.1016/j.engappai.2023.105894

Lancaster University, United Kingdom, Lindley, J. G., Coulton, P., Akmal, H. A., & Pilling, F. L. (2020, September 10). *Signs of the Time: Making AI Legible*. Design Research Society Conference 2020. https://doi.org/10.21606/drs.2020.237

Masotina, M., Musi, E., & Spagnolli, A. (2023). Transparency is Crucial for User-Centered AI, or is it? How this Notion Manifests in the UK Press Coverage of GPT. *Proceedings of the 15th Biannual Conference of the Italian SIGCHI Chapter*, 1–8. https://doi.org/10.1145/3605390.3605413

Moloi, H., Tulloch, N. L., Watkins, D., Perkins, S., Engel, M., Abdullahi, L., Daniels, K., & Zühlke, L. (2022). Understanding the local and international stakeholders in rheumatic heart disease field in Tanzania and Uganda: A systematic stakeholder mapping. *International Journal of Cardiology*, *353*, 119–126. https://doi.org/10.1016/j.ijcard.2022.01.030

Nieder, T. O., Koehler, A., Briken, P., & Eyssel, J. (2020). Mapping key stakeholders' position towards interdisciplinary transgender healthcare: A stakeholder analysis. *Health & Social Care in the Community*, *28*(2), 385–395. https://doi.org/10.1111/hsc.12870

NIST USA. (n.d.). *AI Risk Managment Framework* (NIST AI 100-1).

Rossi, A., & Palmirani, M. (2018). From Words to Images Through Legal Visualization. In U. Pagallo, M. Palmirani, P. Casanovas, G. Sartor, & S. Villata (Eds.), *AI Approaches to the Complexity of Legal Systems* (Vol. 10791, pp. 72–85). Springer International Publishing. https://doi.org/10.1007/978-3-030-00178-0_5

Sangers, T. E., Wakkee, M., Kramer-Noels, E. C., Nijsten, T., & Lugtenberg, M. (2021). Views on mobile health apps for skin cancer screening in the general population: An in-depth qualitative exploration of perceived barriers and facilitators*. *British Journal of Dermatology*, *185*(5), 961–969. https://doi.org/10.1111/bjd.20441

Secinaro, S., Calandra, D., Secinaro, A., Muthurangu, V., & Biancone, P. (2021). The role of artificial intelligence in healthcare: A structured literature review. *BMC Medical Informatics and Decision Making*, *21*(1), 125. https://doi.org/10.1186/s12911-021-01488-9
14

**Appendix**

Table 4. AI act instructions for use related to the capabilities and limitations of the AI system

| The characteristics, capabilities and limitations of performance of the high-risk AI system, including (Art.13.3.b) | | | |
|---|---|---|---|
| **Original text** | **Simplified text** | **Simplified lexicon as per Nist glossary[2] and Wikipedia[3]** | **Icon** |
| Its intended purpose (Art.13.3.b.i). | Intended purpose of the AI system | (no changes) | 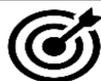 |

---

[2] https://airc.nist.gov/AI_RMF_Knowledge_Base/Glossary

[3] https://en.wikipedia.org/wiki/Logging_(computing)

| Original text | Simplified text | Simplified lexicon as per Nist glossary and Wikipedia | Icon |
|---|---|---|---|
| The level of accuracy, including its metrics, robustness and cybersecurity referred to in Article 15 against which the high-risk AI system has been tested and validated and which can be expected, and any known and foreseeable circumstances that may have an impact on that expected level of accuracy, robustness and cybersecurity (Art.13.3.b.ii). | Accuracy, robustness and cybersecurity. | Correctness, performance resilience and cybersecurity. | 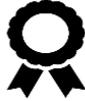 |
| Any known or foreseeable circumstance, related to the use of the high-risk AI system in accordance with its intended purpose or under conditions of reasonably foreseeable misuse, which may lead to risks to the health and safety or fundamental rights referred to in Article 9.2 (Art.13.3.b.iii). | Risks to health, safety, or fundamental rights. | (no changes) | 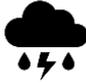 |
| Where applicable, the technical capabilities and characteristics of the high-risk AI system to provide information that is relevant to explain its output (Art.13.3.b.iv). | Capabilities to explain its output. | Capabilities to explain its results. | 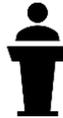 |
| When appropriate, its performance regarding specific persons or groups of persons on which the system is intended to be used (Art.13.3.b.v). | Performance with specific (groups of) persons. | (no changes) | 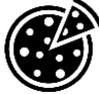 |
| When appropriate, specifications for the input data, or any other relevant information in terms of the training, validation and testing data sets used, taking into account the intended purpose of the high-risk AI system (Art.13.3.b.vi). | Data sets used to train, validate and test the system | (no changes) | 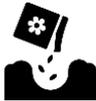 |
| Where applicable, information to enable deployers to interpret the output of the high-risk AI system and use it appropriately (Art.13.3.b.vii). | Appropriate usage | Proper usage | 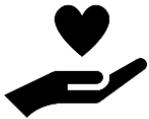 |

| **Other information** | | | |
|---|---|---|---|
| **Original text** | **Simplified text** | **Simplified lexicon as per Nist glossary[4] and Wikipedia[5]** | **Icon** |
| The identity and the contact details of the provider and, where applicable, of its authorised representative; (Art. 13.3.a) | Provider's identity, contact details and authorised representative | Provider's identity, contact details and authorised representative | 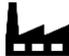 |

---

[4] https://airc.nist.gov/AI_RMF_Knowledge_Base/Glossary

[5] https://en.wikipedia.org/wiki/Logging_(computing)

| Where relevant, a description of the mechanisms included within the high-risk AI system that allows deployers to properly collect, store and interpret the logs in accordance with Article 12. (Art.13.3.f). | Logs' collection, storage and interpretation | Collection, storage and interpretation of activities' record | 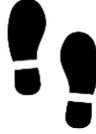 |
|---|---|---|---|
| The computational and hardware resources needed, the expected lifetime of the high-risk AI system and any necessary maintenance and care measures, including their frequency, to ensure the proper functioning of that AI system, including as regards software updates (Art.13.3.e). | Needed computational and hardware resources Expected lifetime Necessary maintenance/care measures, including software updates | Needed computational and hardware resources, maintenance/care and expected lifetime | 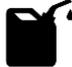 |
| The changes to the high-risk AI system and its performance which have been pre-determined by the provider at the moment of the initial conformity assessment, if any (Art.13.3.c). | Pre-determined changes to AI system | Planned changes to the AI system | 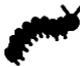 |
| The human oversight measures referred to in Article 14, including the technical measures put in place to facilitate the interpretation of the outputs of the high-risk AI systems by the deployers (Art.13.3.d). | Human oversight measures including facilitation of output interpretation | Human control over the machine, including support in interpreting the results | 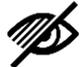 |

**Table 5.** Detail of job positions in the sample

| | |
|---|---|
| managers | |
|     Director (Group Director, Sr. Director, Director) | 7.56% |
|     Manager (Group Manager, Sr. Manager, Manager, Program Manager) | 92.44% |
| health professionals | |
|     Doctor | 62.61% |
|     Nurse | 35.65% |
|     Paramedic | 1.74% |
| patients | |
|     no info | |
| tech | |
|     Account Management, Administration/ Personal Assistant, Chemical / Mechanical / Electrical / Civil Engineering, CX / Customer Experience / Support, Data Analysis, Design or Creative, Healthcare Professional, Engineering (e.g. software), Finance or Accounting, Fundraising, Human Resources, IT / Information Networking / Information Security, Legal, Marketing, Operations, Product or Product Management, Project or Program Management, Public Relations / Communications, Research, Sales / Business Development, Education Professional | 0.87% |
|     Account Management, Administration/ Personal Assistant, Chemical / Mechanical / Electrical / Civil Engineering, CX / Customer Experience / Support, Data Analysis, Design or Creative, Healthcare Professional, Engineering (e.g. software), Finance or Accounting, Fundraising, Human Resources, IT / Information Networking / Information Security, Legal, Sales / Business Development, Product or Product Management | 0.43% |
|     Account Management, Administration/ Personal Assistant, CX / Customer Experience / Support, Design or Creative, Engineering (e.g. software), Finance or Accounting, IT / Information Networking / Information Security, Marketing, Operations, Research, Sales / Business Development | 0.43% |
|     Account Management, Administration/ Personal Assistant, CX / Customer Experience / Support, Engineering (e.g. software), IT / Information Networking / Information Security, Operations, Project or Program Management | 0.43% |
| Account Management, Administration/ Personal Assistant, Design or Creative, Finance or Accounting, Human Resources, IT / Information Networking / Information Security, Operations, Marketing, Product or Product Management, Public Relations / Communications, Research, Sales / Business Development, Education Professional | 0.43% |
| Account Management, Administration/ Personal Assistant, Finance or Accounting, IT / Information Networking / Information Security, Marketing, Operations, Product or Product Management, Project or Program Management, Research, Sales / Business Development, Public Relations / Communications, Design or Creative, Data Analysis, CX / Customer Experience / Support | 0.43% |
| Account Management, CX / Customer Experience / Support, Administration/Personal Assistant, Finance or Accounting, IT / Information Networking / Information Security, Marketing, Product or Product Management, Research, Human Resources, Project or Program Management | 0.43% |
| Account Management, Data Analysis, Finance or Accounting, IT / Information Networking / Information Security, Marketing, Product or Product Management, Public Relations / Communications, Research, Sales / Business Development | 0.43% |
| Account Management, Data Analysis, Healthcare Professional, Engineering (e.g., software), IT / Information Networking / Information Security | 0.43% |
| Account Management, Design or Creative, Engineering (e.g. software), IT / Information Networking / Information Security, Project or Program Management | 0.43% |
| Administration/ Personal Assistant, Account Management, Research, Marketing, Public Relations / Communications, IT / Information Networking / Information Security, Human Resources, Data Analysis | 0.43% |
| Administration/ Personal Assistant, IT / Information Networking / Information Security, Research | 0.43% |
| Chemical / Mechanical / Electrical / Civil Engineering, Engineering (e.g. software), IT / Information Networking / Information Security | 0.43% |
| CX / Customer Experience / Support, Account Management, Data Analysis, Design or Creative, Engineering (e.g. software), IT / Information Networking / Information Security, Marketing, Product or Product Management, Project or Program Management, Public Relations / Communications | 0.43% |
| CX / Customer Experience / Support, Administration/ Personal Assistant, Engineering (e.g. software), IT / Information Networking / Information Security, Research | 0.43% |
| CX / Customer Experience / Support, Data Analysis, IT / Information Networking / Information Security, Project or Program Management | 0.43% |
| CX / Customer Experience / Support, Design or Creative, Engineering (e.g. software), IT / Information Networking / Information Security, Research | 0.43% |
| CX / Customer Experience / Support, IT / Information Networking / Information Security | 0.43% |
| Data Analysis, Administration/ Personal Assistant, Finance or Accounting, IT / Information Networking / Information Security, Project or Program Management, Public Relations / Communications, Sales / Business Development, Account Management | 0.43% |
| Data Analysis, Design or Creative, Engineering (e.g. software), IT / Information Networking / Information Security | 0.43% |
| Data Analysis, Engineering (e.g. software), IT / Information Networking / Information Security | 2.61% |
| Data Analysis, Engineering (e.g. software), IT / Information Networking / Information Security, Chemical / Mechanical / Electrical / Civil Engineering | 0.43% |
| Data Analysis, Engineering (e.g. software), IT / Information Networking / Information Security, Project or Program Management, Product or Product Management | 0.43% |
| Data Analysis, Engineering (e.g. software), IT / Information Networking / Information Security, Research | 0.43% |
| Data Analysis, IT / Information Networking / Information Security | 2.17% |
| Data Analysis, IT / Information Networking / Information Security, Engineering (e.g., software) | 0.43% |
| Data Analysis, IT / Information Networking / Information Security, Operations | 0.43% |
| Design or Creative, Account Management, Administration/ Personal Assistant, Finance or Accounting, IT / Information Networking / Information Security, Research | 0.43% |
| Design or Creative, Engineering (e.g. software), IT / Information Networking / Information Security, Marketing, Product or Product Management, Sales / Business Development, Data Analysis, CX / Customer Experience / Support | 0.43% |
| Design or Creative, IT / Information Networking / Information Security | 1.30% |
| Design or Creative, IT / Information Networking / Information Security, Marketing, Product or Product Management, Project or Program Management, Public Relations / Communications | 0.43% |
| Design or Creative, IT / Information Networking / Information Security, Sales / Business Development | 0.43% |
| Education Professional, IT / Information Networking / Information Security | 0.43% |
| Engineering (e.g. software), Data Analysis, IT / Information Networking / Information Security, Research | 0.43% |
| Engineering (e.g. software), Human Resources, Marketing, Operations, Product or Product Management, Public Relations / Communications, Sales / Business Development, IT / Information Networking / Information Security, Data Analysis, CX / Customer Experience / Support, Account Management, Administration/ Personal Assistant, Design or Creative, Finance or Accounting, Fundraising, Legal | 0.43% |
| Engineering (e.g. software), IT / Information Networking / Information Security | 4.35% |
| Engineering (e.g. software), IT / Information Networking / Information Security, Design or Creative | 0.43% |
| Engineering (e.g. software), IT / Information Networking / Information Security, Design or Creative, Data Analysis, Administration/ Personal Assistant, Account Management | 0.43% |
| Engineering (e.g. software), IT / Information Networking / Information Security, Operations | 0.43% |
| Engineering (e.g. software), IT / Information Networking / Information Security, Operations, Product or Product Management, Project or Program Management, CX / Customer Experience / Support | 0.43% |
| Engineering (e.g. software), IT / Information Networking / Information Security, Product or Product Management, Design or Creative, CX / Customer Experience / Support | 0.43% |
| Engineering (e.g. software), IT / Information Networking / Information Security, Project or Program Management, Product or Product Management | 0.43% |
| Engineering (e.g. software), IT / Information Networking / Information Security, Research | 0.43% |
| Finance or Accounting, IT / Information Networking / Information Security, Data Analysis | 0.43% |
| Human Resources, IT / Information Networking / Information Security | 0.43% |
| Human Resources, IT / Information Networking / Information Security, Data Analysis | 0.43% |
| IT / Information Networking / Information Security | 53.04% |
| IT / Information Networking / Information Security, Administration/ Personal Assistant, Operations | 0.43% |
| IT / Information Networking / Information Security, Administration/ Personal Assistant, Project or Program Management, Legal | 0.43% |
| IT / Information Networking / Information Security, CX / Customer Experience / Support | 0.43% |
| IT / Information Networking / Information Security, CX / Customer Experience / Support, Project or Program Management | 0.43% |
| IT / Information Networking / Information Security, CX / Customer Experience / Support, Research | 0.43% |
| IT / Information Networking / Information Security, Data Analysis | 2.17% |
| IT / Information Networking / Information Security, Data Analysis, Engineering (e.g., software) | 0.43% |
| IT / Information Networking / Information Security, Data Analysis, Engineering (e.g. software), Project or Program Management | 0.43% |
| IT / Information Networking / Information Security, Design or Creative | 0.87% |
| IT / Information Networking / Information Security, Education Professional, Data Analysis, Engineering (e.g., software) | 0.43% |
| IT / Information Networking / Information Security, Engineering (e.g., software) | 1.30% |
| IT / Information Networking / Information Security, Engineering (e.g., software), Education Professional | 0.43% |
| IT / Information Networking / Information Security, Healthcare Professional, Data Analysis, CX / Customer Experience / Support, Public Relations / Communications | 0.43% |
| IT / Information Networking / Information Security, Human Resources | 0.43% |
| IT / Information Networking / Information Security, Legal | 0.43% |
| IT / Information Networking / Information Security, Marketing, Project or Program Management, Engineering (e.g. software), CX / Customer Experience / Support | 0.43% |
| IT / Information Networking / Information Security, Operations | 1.30% |
| IT / Information Networking / Information Security, Operations, Data Analysis, Design, or Creative | 0.43% |
| IT / Information Networking / Information Security, Operations, Research, Administration/ Personal Assistant | 0.43% |
| IT / Information Networking / Information Security, Product or Product Management | 0.43% |
| IT / Information Networking / Information Security, Product or Product Management, Sales / Business Development | 0.43% |
| IT / Information Networking / Information Security, Project or Program Management | 0.43% |
| IT / Information Networking / Information Security, Research, Data Analysis | 0.43% |
| Legal, Finance or Accounting, IT / Information Networking / Information Security, Marketing, Data Analysis | 0.43% |
| Marketing, Finance, or Accounting, IT / Information Networking / Information Security | 0.43% |
| Operations, Project or Program Management, IT / Information Networking / Information Security, Finance or Accounting | 0.43% |
| Product or Product Management, Public Relations / Communications, Sales / Business Development, IT / Information Networking / Information Security, Administration/ Personal Assistant | 0.43% |
| Project or Program Management, IT / Information Networking / Information Security | 0.43% |
| Public Relations / Communications, IT / Information Networking / Information Security | 0.43% |
| Research, IT / Information Networking / Information Security, Engineering (e.g., software), Data Analysis | 0.43% |
| Research, Operations, IT / Information Networking / Information Security | 0.43% |
| Sales / Business Development, Product or Product Management, Project or Program Management, IT / Information Networking / Information Security, Engineering (e.g. software), CX / Customer Experience / Support, Operations | 0.43% |

**Table 6.** Words in the IFU section titles marked as obscure by the respondents

| word | frequency |
|---|---|
| resilience | 203 |
| computational | 144 |
| authorized representative | 83 |
| performance | 66 |
| fundamental rights | 63 |
| datasets | 62 |
| activity record | 54 |
| proper usage | 41 |
| correctness | 38 |
| interpretation/interpreting | 31 |
| human control | 30 |
| planned changes | 30 |
| cybersecurity | 27 |
| expected lifetime | 25 |
| provider's | 24 |
| risks | 24 |
| identity | 23 |
| validate | 21 |
| AI system | 16 |
| capabilities | 16 |
| intended | 15 |
| specific | 15 |
| health | 14 |
| persons | 14 |
| storage | 14 |
| contact details | 13 |
| explain | 13 |
| hardware | 13 |
| collection | 12 |
| results | 11 |
| purpose | 10 |
| maintenance/care | 9 |
| safety | 9 |
| train | 9 |
| groups | 7 |
| limitations | 6 |
| machine | 6 |
| resources | 6 |
| test | 6 |
| over | 4 |
| support | 4 |
| used | 1 |

21